\pdfoutput=1
\documentclass[prx,superscriptaddress,amsmath,amssymb,twocolumn,nofootinbib]{revtex4-2}
\usepackage{graphicx,bm,amsmath}
\usepackage[usenames,dvipsnames]{xcolor}
\usepackage[colorlinks,bookmarks=false,citecolor=NavyBlue,linkcolor=Red,urlcolor=blue,
]{hyperref}
\usepackage[percent]{overpic}
\usepackage[bbgreekl]{mathbbol}
\usepackage{xcolor}
\usepackage[normalem]{ulem}
\usepackage{algorithm}
\usepackage{braket}
\usepackage{algpseudocode}
\usepackage{comment}
\usepackage{amsthm}
\usepackage{enumerate}
\usepackage{lipsum}
\def\doi{http://dx.doi.org/}
\newcommand{\be}{\begin{equation}}
\newcommand{\ee}{\end{equation}}
\newcommand{\bec}{\begin{equation*}}
\newcommand{\eec}{\end{equation*}}
\newcommand{\bea}{\begin{eqnarray}}
\newcommand{\eea}{\end{eqnarray}}

\newcommand{\tr}{\text{tr}}   

\usepackage{color}

\newcommand{\titleinfo}{Entanglement Rényi Negativity across the Finite-Temperature Transition in the O(3) Universality Class}
\begin{document}
\title{\titleinfo}
\author{Dong-Xu Liu}
\affiliation{Department of Physics, School of Science and Research Center for Industries of the Future, Westlake University, Hangzhou 310030,  China}
\affiliation{Institute of Natural Sciences, Westlake Institute for Advanced Study, Hangzhou 310024, China}

\author{Yi-Ming Ding}
\affiliation{Department of Physics, School of Science and Research Center for Industries of the Future, Westlake University, Hangzhou 310030,  China}
\affiliation{Institute of Natural Sciences, Westlake Institute for Advanced Study, Hangzhou 310024, China}
\affiliation{State Key Laboratory of Surface Physics and Department of Physics, Fudan University, Shanghai 200438, China}

\author{Zhe Wang}
\email{wangzhe90@westlake.edu.cn}
\affiliation{Department of Physics, School of Science and Research Center for Industries of the Future, Westlake University, Hangzhou 310030,  China}
\affiliation{Institute of Natural Sciences, Westlake Institute for Advanced Study, Hangzhou 310024, China}

\author{Zheng Yan}
\email{zhengyan@westlake.edu.cn}
\affiliation{Department of Physics, School of Science and Research Center for Industries of the Future, Westlake University, Hangzhou 310030,  China}
\affiliation{Institute of Natural Sciences, Westlake Institute for Advanced Study, Hangzhou 310024, China}
    
\begin{abstract}
The fate of quantum entanglement at finite-temperature phase transitions remains an open question, particularly for continuous symmetry breaking where zero-temperature Goldstone modes generate long-range correlations. 
Using large-scale quantum Monte Carlo simulations, we investigate the third R\'enyi negativity across the $O(3)$ transition in the three-dimensional Heisenberg antiferromagnet—the first such study for a thermal critical point with continuous symmetry.
We uncover two fundamental results. First, the negativity exhibits a pure area law at the critical point, with the subleading constant term vanishing within statistical uncertainty. 
This demonstrates that thermal fluctuations completely destroy the long-range entanglement present at zero temperature—the divergent classical correlation length leaves no imprint on quantum entanglement itself.
Second, despite this absence of singular behavior in the negativity, its temperature derivative follows the exact scaling of the specific heat, yielding critical exponents $-\alpha/\nu = 0.190(1)$ and $1/\nu = 1.350(5)$ in precise agreement with the $O(3)$ universality class. 
Our work establishes that while quantum entanglement is blind to thermal criticality, its thermodynamic derivatives encode the full universal scaling, revealing an unexpected connection between entanglement and classical phase transitions. Furthermore, this also provides further evidence that R\'enyi negativity can still effectively shield classical correlations in systems with continuous symmetry.

\end{abstract}

\maketitle
\section{Introduction}
The study of many-body critical phenomena has been significantly advanced through the application of quantum information measures. 
For bipartite entanglement in general mixed states, entanglement negativity, rooted in the positive partial transpose criterion, has emerged as a key computable probe for this purpose \cite{Horodecki1997Separability,Asher1996Separability,Hassan2017Partial,vidal2002computable, plenio2005logarithmic}. 
However, the direct calculation of negativity is generally challenging in many-body systems. 
Therefore, the Rényi negativity has been introduced as a more tractable proxy for characterizing mixed-state entanglement that can be accessed both analytically and numerically~\cite{Wu2020Entanglement,Chung2014Entanglement,Ding2025Tracking,
calabrese2012entanglement, Calabrese_2013_replicaTrick,calabrese2013entanglement,chuang2014negmoment_qmc,Alba2013negativity,Neven2021p3ppt,fang2025qmcnegativity, Wang2025negativity, wang2025untwistednegativity}. 
It is defined as
\begin{equation}\label{eq:def}
R_n = -\ln\!\left(\frac{\tr[(\rho^{T_B})^n]}{\tr[\rho^n]}\right),
\end{equation}
where $\rho$ is the density matrix and $T_B$ denotes the partial transpose with respect to subsystem $B$.

For quantum phase transitions, similar to entanglement entropy, Rényi negativity obeys an area law, scaling with the size of the entangling boundary, i.e., $R_n\sim L^{d-1}$~\cite{Hart2018Entanglement,Sherman2016Nonzerotemperature}. 
Subleading corrections to this leading behavior can usually encode universal information about the underlying quantum correlations, including long-range entanglement~\cite{Lee2013Entanglement, Castelnovo2013Negativity,Wen2016Topological}.
Although enormous progress has been made over the past two decades in understanding the entanglement properties of pure quantum states~\cite{metlitski2011entanglement,wang2025bipartite,Kulchytskyy2015Detecting,Nicolas2016Quantum,Metlitski2009Entanglement}, the study of entanglement in interacting many-body quantum systems in mixed states remains comparatively less developed, even for equilibrium Gibbs states~\cite{Shapourian2019Finite,Fan2024Diagnostics,Wybo2020Entanglement,Lu2020Detecting}.

For a purely classical Hamiltonian (e.g., the 3D classical Ising model) at its finite-temperature critical point, the equilibrium density matrix is diagonal in a product basis, and therefore contains no quantum entanglement for any bipartition. 
By contrast, for quantum Hamiltonians such as the 2D transverse-field Ising model (TFIM), recent studies have shown that the entanglement negativity can remain finite over a broad temperature range around the finite-temperature critical point~\cite{Wu2020Entanglement,Lu2019Singularity,Javanmard2018Sharp}.
However, these results primarily concern systems with discrete symmetry breaking. 
For thermal phase transitions associated with continuous symmetry breaking, the ordered phase supports massless  Goldstone modes, which can strongly affect low-energy correlations and thermodynamic properties. 
At present, it remains unclear to what extent such modes influence mixed-state entanglement at finite temperature, and in particular whether they qualitatively modify the behavior of Rényi negativity near a thermal critical point.

Furthermore, previous studies have suggested that, at finite-temperature phase transitions, 
the temperature derivative of the area-law coefficient of the negativity (and its Rényi counterparts) 
captures singular contributions arising from long-range correlations, and is expected to scale in the same way as the specific heat~\cite{Lu2019Singularity,Wu2020Entanglement}.
In particular, the area-law coefficient of the Rényi negativity becomes singular across the transition, 
reflecting the underlying thermodynamic criticality.
This scaling relation has been verified for the special case where the specific-heat critical exponent 
$\alpha = 0$, such as in the two-dimensional Ising universality class, where the singular part of the 
specific heat diverges logarithmically with system size~\cite{Ferdinand1969Bounded,AuYang1975Bounded}. 
However, numerical evidence for this relation in systems with $\alpha \neq 0$ remains limited. From another perspective, how the divergent non-quantum correlation length would specifically manifest in R\'enyi negativity still lacks sufficient evidences and remains understudied. More quantitative questions need to be further explored.

In this work, we present an extensive numerical study of the R\'enyi negativity in the 3D Heisenberg antiferromagnet on a cubic lattice using quantum Monte Carlo (QMC) simulations. 
In contrast to the two-dimensional transverse-field Ising model studied in previous works, whose finite-temperature transition is associated with discrete $\mathbb{Z}_2$ symmetry breaking, the 3D Heisenberg antiferromagnet undergoes a thermal phase transition with continuous $O(3)$ symmetry breaking. 
The ordered phase therefore supports massless  Goldstone modes which lead long-range entanglement contribution in entanglement entropy at zero temperature, providing a natural setting to examine how such collective excitations influence mixed-state entanglement near a thermal critical point. 
Moreover, the corresponding universality class is characterized by a nonzero (in fact negative) specific-heat critical exponent $\alpha$, which allows us to test the proposed scaling relation between the temperature derivative of the negativity area-law coefficient and the specific heat beyond the special case $\alpha=0$. 
This model therefore provides an ideal platform to address the two questions raised above.

\begin{figure}[ht!]
\centering
\includegraphics[width=1.0\linewidth]{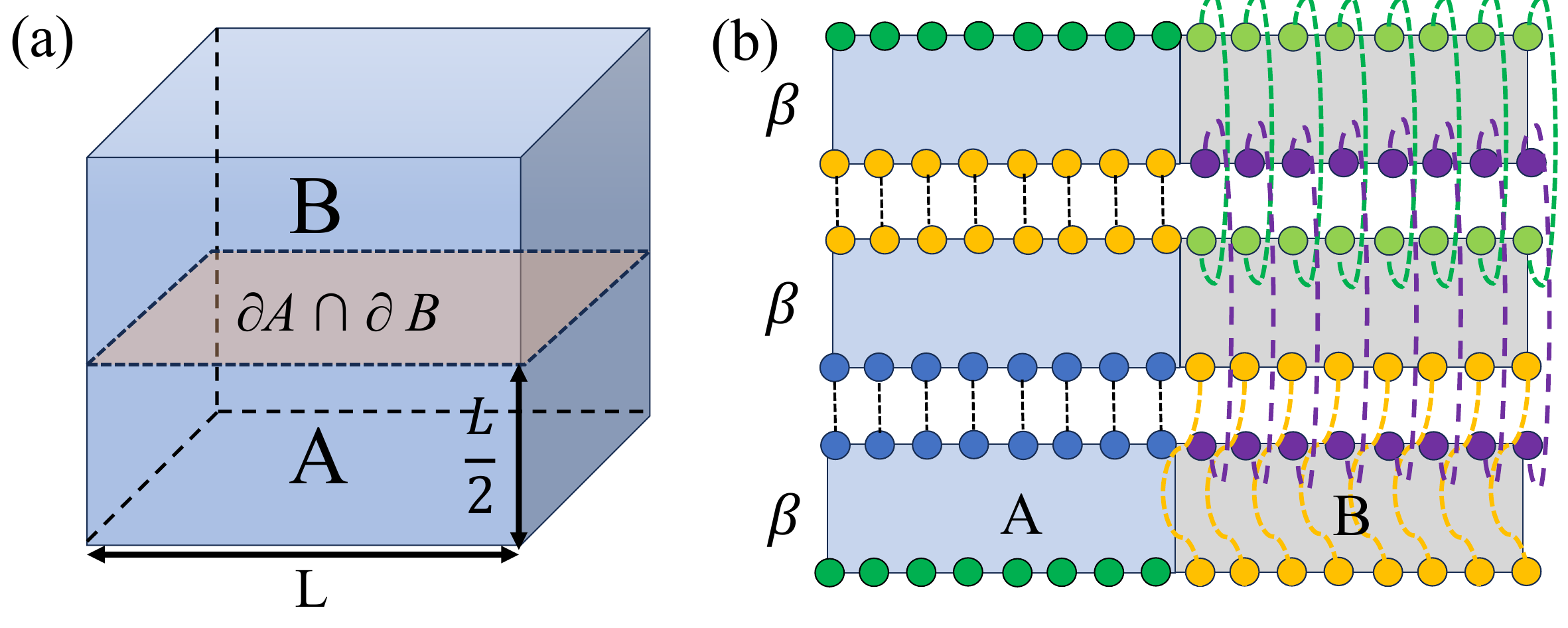}
\caption{(a) Schematic of the partitioned subsystems A and B on a three-dimensional cubic lattice, sharing an interface $\partial A \cap \partial B$. (b) The space-time manifold in path integral used to compute the Rényi negativity $R_{3}$. The horizontal axis denotes the spatial spin direction and the vertical axis corresponds to imaginary time. Spins
of the same color are connected along the imaginary time direction.}
\label{fig:neg_def_cut}
\end{figure}

\section{Model and Method}
We consider the 3D spin-1/2 Heisenberg antiferromagnet on a cubic lattice of size $L\times L\times L$ with periodic boundary conditions. 
The Hamiltonian of the system is given by
\begin{equation}
\begin{split}
H= & J \sum_{\langle ij\rangle} \mathbf{S}_i\cdot \mathbf{S}_j,
\end{split}
\label{eq:sys_ham}
\end{equation}
where $J>0$ denotes the antiferromagnetic exchange interaction between nearest-neighbor spins and the summation runs over all nearest-neighbor pairs $\langle ij\rangle$ of the cubic lattice.
The positive exchange coupling favors antiparallel alignment of neighboring spins. 
On the bipartite cubic lattice, this leads to long-range N\'eel order in the ground state, characterized by a staggered magnetization that spontaneously breaks the global SU(2) spin-rotation symmetry. 
At finite temperature, this ordered phase is separated from the high-temperature paramagnetic phase by a continuous phase transition at $T_c \approx 0.938$ (in units of $J=1$)~\cite{Sandvik1998Critical,Xiaoling2008Heisenberg}.

A key feature of this model is that the finite-temperature phase transition from the N\'eel state to the paramagnet belongs to the three-dimensional $O(3)$ universality class, which is same to the quantum phase transition in 2D quantum systems at zero temperature. 
Near the critical point, the long-wavelength fluctuations of the staggered magnetization are effectively described by a three-dimensional classical field theory with $O(3)$ symmetry. 
Consequently, the critical behavior is characterized by universal critical exponents. 
In particular, the correlation-length exponent is $\nu \approx 0.7116$, and the specific-heat exponent is $\alpha \approx -0.1336$, consistent with the hyperscaling relation $\alpha = 2-d\nu$~\cite{Xiaoling2008Heisenberg,PhysRevB.21.3976,Guillou_JDP_1989}. 
These exponents govern the singular behavior near the critical temperature $T_c$, where the correlation length diverges as $\xi \sim |T-T_c|^{-\nu}$.

The R\'enyi negativity provides a computable measure of quantum entanglement for mixed states and is particularly useful for diagnosing bipartite entanglement in finite-temperature systems. 
For a bipartition without geometric singularities—such as a smooth boundary or a straight cut dividing the system into two extended regions without corners, as illustrated in Fig.~\ref{fig:neg_def_cut}(a)the leading contribution to the R\'enyi negativity at criticality typically follows an area law,
\begin{equation}
R_n = a \, |\partial A \cap \partial B| + \gamma ,
\end{equation}
where $|\partial A \cap \partial B|$ denotes the area of the interface between the two subsystems~\cite{Lu2020Structure,Grover2011Entanglement}. 
The coefficient $a$ is nonuniversal, while $\gamma$ represents a subleading constant contribution that usually encodes long-range quantum entanglement. 
At a conventional finite-temperature critical point in systems such as the Heisenberg model, the transition is governed by thermal fluctuations. 
While the classical correlation length diverges, quantum correlations are not expected to develop long-range coherence. 
Consequently, the subleading constant contribution $\gamma$ is generally expected to vanish in the thermodynamic limit, and the singular behavior of $R_n$ near the transition, if exist, is dominated by the divergence of the classical correlation length~\cite{Wen2016Topological,Lu2020Detecting,Lu2019Singularity}. 

To investigate the critical behavior of entanglement at finite temperature, we study the R\'enyi negativity in the three-dimensional antiferromagnetic Heisenberg model. Using quantum Monte Carlo simulations, we analyze the behavior of $R_n$ with $n=3$ near the finite-temperature phase transition. Specifically, we employ quantum Monte Carlo with the reweight-annealing method~\cite{ding2024reweight,Ding2025Tracking,Ding2025Evaluating,wang2025bipartite,wang2026addressing} to compute $R_3$, the space-time manifold in path integral is shown in the Fig.\ref{fig:neg_def_cut} (b). Further details of this method are provided in the Appendix \ref{app}.

\begin{figure}[ht!]
\centering
\includegraphics[width=0.8\linewidth]{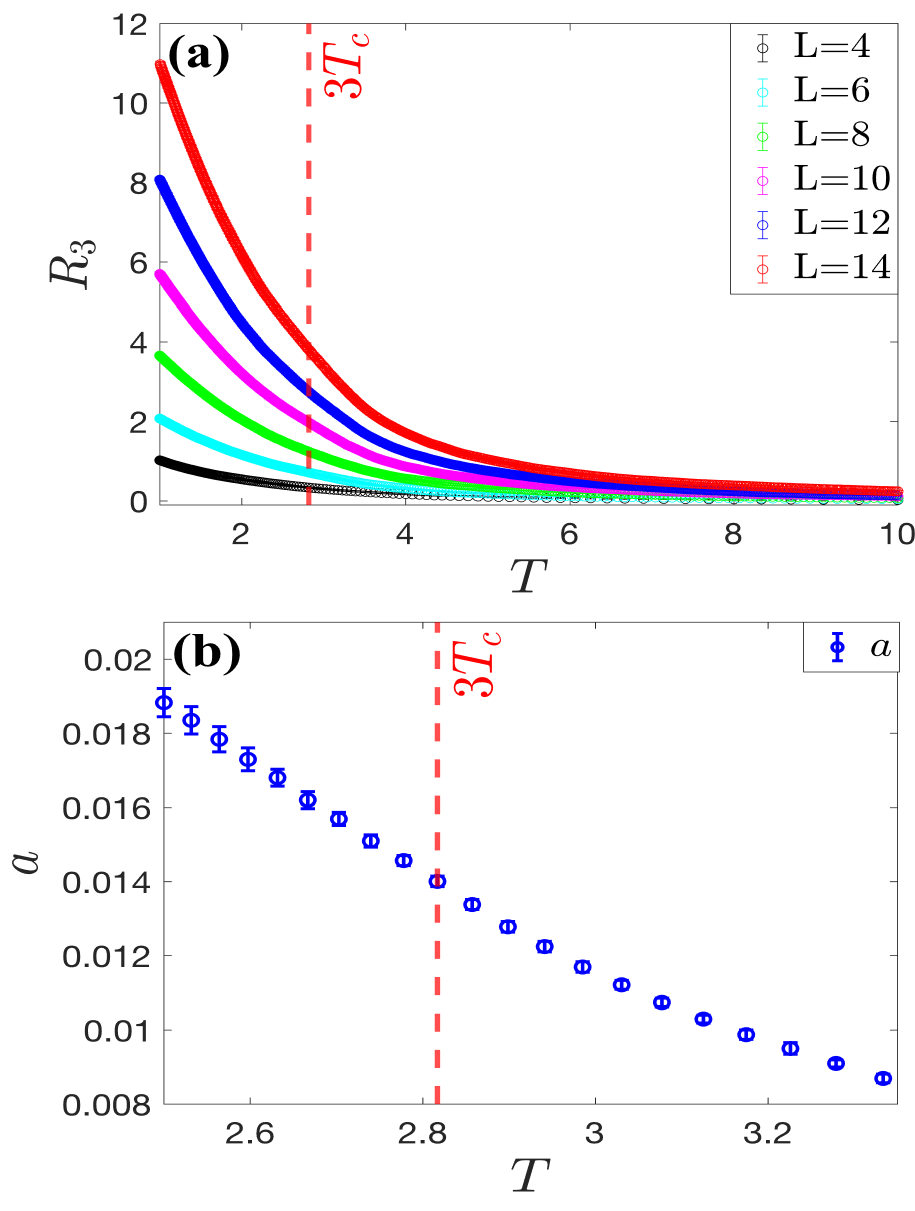}
\caption{The Rényi negativity $R_3$ was computed for the three-dimensional Heisenberg model and fitted to the area law $R_3 = aL^2 + \gamma$. (a) $R_3$ as a function of temperature $T$.(b) Temperature dependence of the fitted coefficient $a$.}
\label{fig:neg_3Tc_a}
\end{figure}

\section{Results}
The first question is whether the curve of $R_3$ $vs$ $T$ displays any singular behaviors. In a (2+1)D Heisenberg model, its quantum phase transition is also 3D $O(3)$ universality. The previous studies \cite{wang2025bipartite,wang2025universal} show that the entanglement entropy has variations in concavity and convexity near the quantum phase transition point from a N\'eel order to a disordered phase.
Similarly, we now explore the negativity behaviors from N\'eel order to disordered phase in the 3D Heisenberg model with a finite temperature 3D $O(3)$ phase transition.
As shown in Fig.~\ref{fig:neg_3Tc_a}(a), the third-order Rényi negativity $R_3$ is computed as a function of the temperature $T$. At the critical point, $R_3$ itself does not display any clear singular behavior or pronounced nonanalytic feature. 
However, previous studies have suggested that singularities in the third Rényi negativity may appear around $\beta/3$, rather than at $\beta$ corresponding to the actual thermal transition~\cite{Wu2020Entanglement}. 
This shift arises from the replica structure of the Rényi negativity, which effectively rescales the temperature in the replicated manifold.
Additionally, as shown in Fig.~\ref{fig:neg_3Tc_a}(b), the coefficient $a$ of the area law $R_3=aL^2+\gamma$ exhibits no clear singular behavior across $3T_c$, indicating that the area-law contribution evolves smoothly near this point. This is totally different from the behaviors of entanglement entropy in (2+1)D $O(3)$ quantum phase transition \cite{helmes2014entanglement}, in which the coefficient $a$ peaks at the critical point.
Therefore, the singularity introduced by the divergent classical correlation length, at least, seemingly never comes into the leading coefficient, even for the phase transition of continuous symmetry.

\begin{figure}[ht!]
\centering
\includegraphics[width=0.8\linewidth]{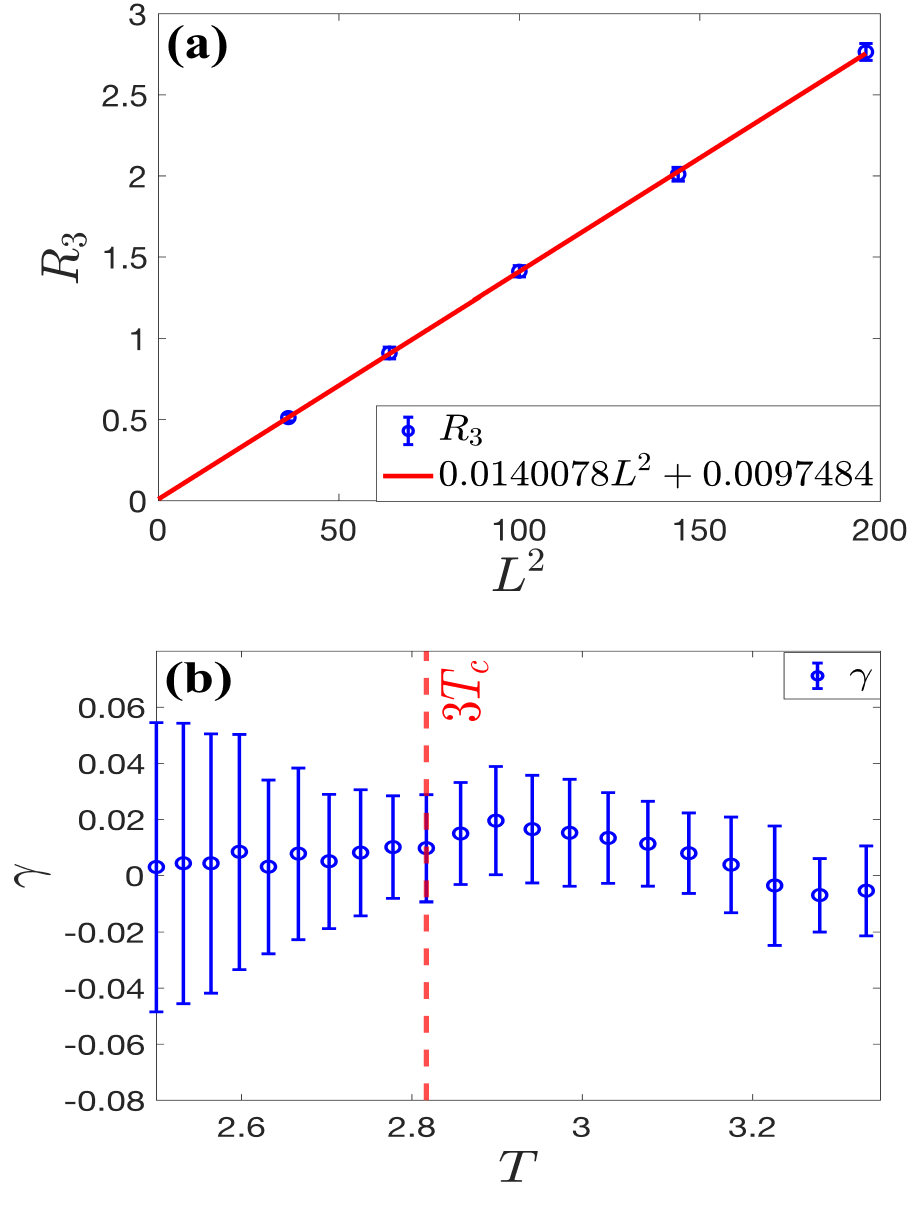}
\caption{The Rényi negativity $R_3$ was computed for the three-dimensional Heisenberg model. (a) $R_3$ follows the area law $R_3 = aL^2 + \gamma$ at the rescaled temperature $3T_c$. (b) The temperature dependence of the fitted subleading coefficient $\gamma$ near $3T_c$, showing that it is consistent with zero within the statistical uncertainty.}
\label{fig:neg_3Tc_b}
\end{figure}

To further investigate the scaling properties, we analyzed the behavior of $R_3$ near the rescaled temperature $3T_c$ by fitting it to the area law $R_3 = aL^2 + \gamma$, where $L^2$ is the interface area between subsystems $A$ and $B$ in the bipartite geometry. System sizes $L = 4, 6, \dots, 14$ were used. 
As shown in Fig.~\ref{fig:neg_3Tc_b}(a), $R_3$ is plotted as a function of the interface area $L^2$ at $T = 3T_c$. The data are well described by a linear fit $R_3 = aL^2 + \gamma$, indicating area-law scaling of the Rényi negativity. It shows that there is no extra nonzero constant correction in the scaling, i.e. $\gamma=0$, which points to no long-range entanglement. Oppositely, the entanglement entropy scales as $S =a\frac{L^{d-1}}{a^{d-1}} + \gamma\frac{L^{d-1}}{\xi^{d-1}}$ with $\gamma \neq 0$ in $O(N)$ quantum phase transitions~\cite{Metlitski2009Entanglement,wang2026universal}, where the divergent correlation length $\xi$ contributes to the constant term. It reflects that the R\'enyi negativity displays a a good shielding of the divergent classic correlations at the critical point.

We keep trace to the nearby phases, especially the N\'eel order with Goldstone modes.
Figure~\ref{fig:neg_3Tc_b}(b) shows the behavior of $\gamma$ in the vicinity of the rescaled transition temperature $3T_c$. Within statistical uncertainties, the constant term $\gamma$ is consistent with zero. This observation suggests that quantum correlations remain short-ranged across a conventional finite-temperature phase transition. The R\'enyi negativity can faithfully describe the entanglement property even for a continuous-symmetry-breaking phase.

Although the various behaviors of the $R_3$ point to no singularity at the (rescaled) phase transition point, we believe that the critical information should be still encoded in the space-time manifold of path integral, because the R\'enyi negativity can be treated as a difference of free energies of different space-time manifolds and the difference only exists on the boundary \cite{Lu2019Singularity}.
To probe possible hidden singular behavior in $R_3$, we numerically compute $dR_3/dT$ in the vicinity of $3T_c$ using a finite-difference method. 
It can be shown that the derivative of $R_3$ with respect to temperature, normalized by the boundary area $|\partial A \cap \partial B|$, is equivalent to the derivative of the area-law coefficient $a$, namely $dR_3/dT \propto da/dT$. 
Previous works \cite{Wu2020Entanglement,Lu2019Singularity} argued that $da/dT$ exhibits the same scaling behavior as the specific heat,
\begin{equation}
    \frac{da}{dT} \sim a'_0 + C_s ,
\end{equation}
where $a'_0$ denotes the nonsingular contribution and $C_s$ represents the singular part of the specific heat~\cite{Wu2020Entanglement}. 
This relation has been verified for the two-dimensional Ising universality class, where the specific-heat critical exponent $\alpha=0$ and $C_s \sim \ln L$. 
For systems with $\alpha \neq 0$, the singular part instead follows the scaling form $C_s \sim |t|^{-\alpha}$. 
To date, numerical evidence for this scaling behavior in such cases remains limited.



\begin{figure}[ht!]
\centering
\includegraphics[width=1.0\linewidth]{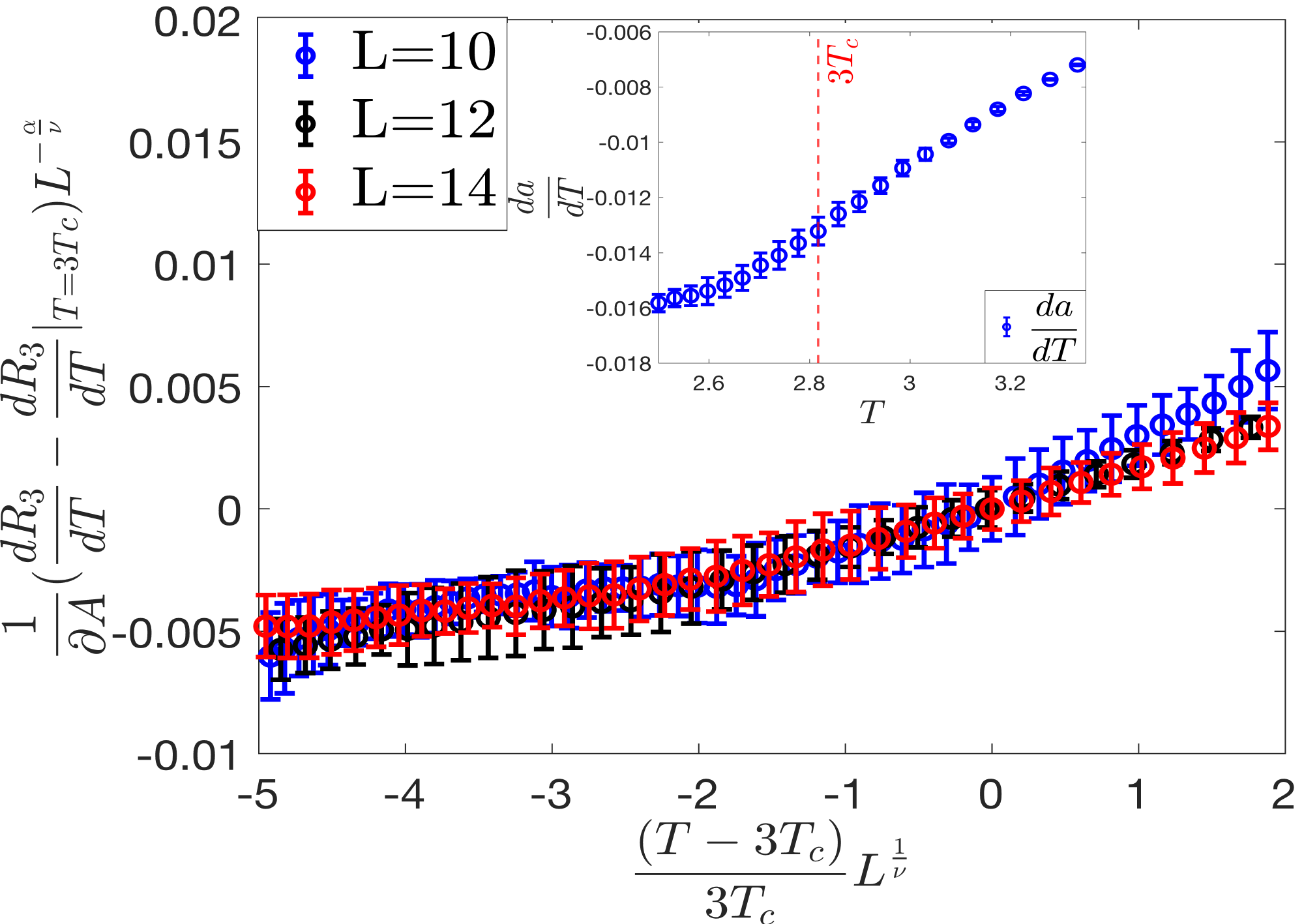}
\caption{Finite-size scaling collapse of data for system sizes $L=10,12,14$ using Eq.(\ref{eq:data_collaps}). The horizontal axis is the scaled variable $tL^{1/\nu}$, where $t=(T-3T_{c})/3T_{c}$ is the reduced temperature. The vertical axis represents the singular part of the interface density of the derivative, scaled as $(1/\partial A) [(dR_{3}/dT)-(dR_{3}/dT)\vert_{T=3T_{c}}]L^{-\alpha/\nu}$, which captures the critical behavior near $3T_{c}$. Inset: The behavior of the derivative of the area-law coefficient near the rescaled temperature $3T_{c}$.} 
\label{fig:dneg_collaps_3Tc}
\end{figure}

To suppress nonsingular contributions while isolating the singular part of the Rényi negativity, we consider the difference relative to the critical point,
\begin{equation}
\begin{split}
\frac{1}{L^{d-1}}
\left(
\frac{dR_3}{dT}
-
\left.\frac{dR_3}{dT}\right|_{T=3T_c}
\right)
&=
a_0'(T)-a_0'(3T_c) \\
&\quad + b|T-3T_c|^{-\alpha}.
\end{split}
\label{eq:iso_sing_term}
\end{equation}
The first term represents the analytic background and, to leading order, scales linearly with $T-3T_c$. 
Close to the critical point, the nonanalytic contribution associated with the universal term therefore dominates.

In a finite system, the divergence of the correlation length $\xi \sim |t|^{-\nu}$ is effectively cut off by the system size $L$. 
Consequently, the singular part obeys the finite-size scaling form
\begin{equation}
\frac{1}{L^{d-1}}
\left(
\frac{dR_3}{dT}
-
\left.\frac{dR_3}{dT}\right|_{T=3T_c}
\right)
=
L^{\alpha/\nu}\,F\!\left(tL^{1/\nu}\right).
\label{eq:data_collaps}
\end{equation}
where $t=(T-3T_c)/3T_c$ is the reduced temperature, $\nu$ is the correlation-length critical exponent, and $F$ is a dimensionless scaling function.

We perform a finite-size scaling analysis based on Eq.~(\ref{eq:data_collaps}) using data for system sizes $L=10,12,$ and $14$. 
As shown in Fig.~\ref{fig:dneg_collaps_3Tc}, the data collapse well onto a single universal curve. 
Fitting the scaling function yields critical exponents $-\alpha/\nu \approx 0.190$ and $1/\nu \approx 1.350$, which are consistent with the established values for the three-dimensional $O(3)$ universality class ($-\alpha/\nu = 0.188$ and $1/\nu = 1.406$) \cite{Xiaoling2008Heisenberg,PhysRevB.21.3976,Guillou_JDP_1989}. 
This agreement supports the scaling form proposed in Eq.~(\ref{eq:data_collaps}), indicating that the singular part of the temperature derivative of the Rényi negativity follows the same critical scaling behavior as the specific heat. Furthermore, as shown in the inset of Fig.~\ref{fig:dneg_collaps_3Tc}, $da/dT$ does not exhibit the divergent behavior observed in the 2D TFIM~\cite{Wu2020Entanglement} at its finite-temperature transition. This absence of divergence stems from the fact that the critical exponent ratio $\alpha/\nu$ is negative in this model.


 \section{Conclusions}
In this work, we have systematically investigated the critical behavior of the third Rényi negativity $R_{3}$ across the finite-temperature phase transition of the three-dimensional Heisenberg model using large-scale quantum Monte Carlo simulations. 
At this temperature, the negativity exhibits an area-law scaling, $R_3 = aL^2 + \gamma$. 
Within statistical uncertainties, the subleading constant term $\gamma$ is consistent with zero, indicating that no additional contribution associated with long-range quantum entanglement is detected. This behavior reflects the distinct nature of finite-temperature criticality with divergent classical correlation length. 
It strongly supports that the R\'enyi negativity will not mistakenly include the contributions of the classic correlations in the entanglement behaviors, even in a phase transition of continuous symmetry.

A key observation of this work is the striking absence of any divergent singularity in either $R_{3}$ or its temperature derivative $d R_{3}/dT$ in the vicinity of $3T_{c}$. While this smooth behavior appears to contrast with the logarithmic divergence typically found in the Ising universality class, it in fact reflects the distinct critical scaling of the O(3) model. To elucidate the origin of this behavior, we performed a detailed scaling analysis. We show that, due to the critical exponents of the 3D O(3) universality class, specifically $-\alpha > 0$, the non-analytic part of $d R_{3}/dT$ vanishes continuously at the transition, rather than diverging. Nevertheless, its critical behavior remains governed by the scaling form of the specific heat. In particular, an explicit scaling relation has been demonstrated that the interfacial density of 
$d R_{3}/dT$ on the boundary $\vert \partial A \cap \partial B \vert$ shares the same singular scaling form as the specific heat. 
This study fills the gap in the research on entanglement in the mixed-state's continuous symmetry-breaking phase transition.
\section{Acknowledgements}
 ZW is supported by the China Postdoctoral Science Foundation under Grants No.2024M752898. 
The work is supported by the Scientific Research Project (No.WU2025B011) and the Start-up Funding of Westlake University.
The authors thank the IT service office and the high-performance computing center of Westlake University.

\bibliographystyle{apsrev4-1}
\bibliography{bibliography}

\clearpage
\appendix
\setcounter{equation}{0}
\setcounter{figure}{0}
\renewcommand{\theequation}{S\arabic{equation}}
\renewcommand{\thefigure}{S\arabic{figure}}
\setcounter{page}{1}
\begin{widetext}
	

\section{Quantum Monte Carlo with Reweight-Annealing}
\label{app}
The central idea of calculating the Rényi negativity via a reweighting–annealing approach is to use quantum Monte Carlo (QMC) simulations to estimate the ratio between two partition functions~\cite{Neal2001,PhysRevB.110.165152,wang2025}. 
For clarity, we first introduce two generalized partition functions,
\begin{equation}
Z_n = \mathrm{Tr}\!\left[\rho^n\right], 
\qquad 
Z_n^{T_B} = \mathrm{Tr}\!\left[(\rho^{T_B})^n\right],
\end{equation}
Here the density matrix is given by $\rho = e^{-\beta H}/Z$, with $Z = \mathrm{Tr}[e^{-\beta H}]$ the usual thermal partition function.

Based on the definition in Eq.~(\ref{eq:def}), the Rényi negativity can be written as
\begin{equation}
R_n = \ln \frac{Z_n^{T_B}}{Z_n},
\end{equation}
which reduces the problem of computing entanglement to evaluating the ratio between two partition functions.
\begin{figure}[ht!]
\centering
\includegraphics[width=0.75\linewidth]{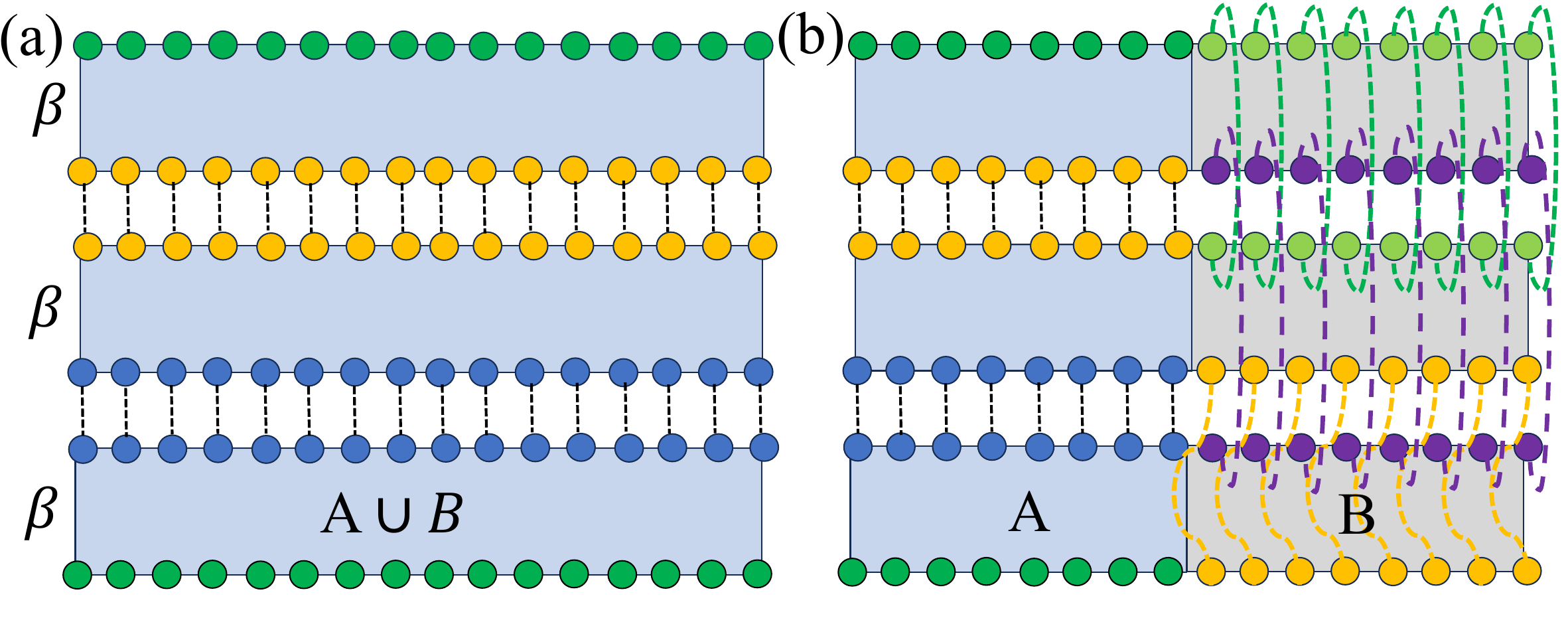}
\caption{The horizontal axis denotes the spatial spin direction, while the vertical axis corresponds to imaginary time. Spins of the same color are connected along the imaginary time direction. (a) Manifold representing $Z_{3}$, where three replicas are connected along the imaginary time direction for the entire system $A \cup B$. (b) The twisted manifold corresponding to $Z^{T_{B}}_{3}$, in which the three replicas are connected within subsystem $A$, while in subsystem $B$, the connections are no longer cyclic due to the partial transpose operation $T_{B}$.}
\label{fig:neg_def_manifold}
\end{figure}
In the imaginary-time path integral formulation, subsystem $B$ in partition function $Z_n$ is defined on a replica-disconnected manifold parameterized,  as shown in Fig.~\ref{fig:neg_def_manifold}(a), by a continuous parameter $\lambda$ (the annealing parameter), which may correspond to a physical parameter such as the inverse temperature $\beta$. 
In contrast, subsystem $B$ in $Z_n^{T_B}$ has a similar form but is defined on a twisted manifold with partial-transpose boundary conditions~\cite{Wen2016Topological,Chung2014Entanglement,Ding2025Tracking,Ding2025Evaluating}.

To probe the parameter dependence, we consider the difference
\begin{equation}
    \Delta R_n = R_n(\lambda'') - R_n(\lambda'),
\end{equation}
which represents the change in the Rényi negativity between two parameter points $\lambda'$ and $\lambda''$. 
Using the definition above, this quantity can be written as
\begin{equation}
\Delta R_n
=
\ln \frac{Z_n^{T_B}(\lambda'')}{Z_n^{T_B}(\lambda')}
-
\ln \frac{Z_n(\lambda'')}{Z_n(\lambda')}.
\label{eq:diff_neg}
\end{equation}
For the ratios $Z_n^{T_B}(\lambda'')/Z_n^{T_B}(\lambda')$ and 
$Z_n(\lambda'')/Z_n(\lambda')$ appearing in Eq.~(\ref{eq:diff_neg}), 
both can be evaluated using reweighting or annealing techniques. 

Specifically, to estimate the ratios, we write the partition function as
\begin{equation}
    Z_n(\lambda')=\sum_{\gamma} W_{\gamma}(\lambda'),
\end{equation}
where $W_{\gamma}(\lambda')$ denotes the statistical weight of configuration $\gamma$, and the sum runs over all configurations in the configuration space. 
The ratio of partition functions at two different parameter values can then be expressed as 
\begin{equation}
    \frac{Z_n(\lambda'')}{Z_n(\lambda')}
=
\left\langle
\frac{W_{\gamma}(\lambda'')}{W_{\gamma}(\lambda')}
\right\rangle_{Z_n(\lambda')},
\end{equation}
where $\langle\cdots\rangle_{Z_n(\lambda')}$ denotes the ensemble average over configurations sampled from the probability distribution defined by $Z_n(\lambda')$. 
An analogous expression applies to the ratio involving $Z_n^{T_B}$.

To compute the ratio $Z_n(\lambda'')/Z_n(\lambda')$ 
(the same procedure applies to the ratio involving $Z_n^{T_B}$), 
one may directly evaluate the reweighting estimator introduced above. 
However, this estimator relies on the overlap between the probability distributions corresponding to the two partition functions and can become exponentially small when the parameters $\lambda'$ and $\lambda''$ are far apart, making accurate sampling difficult in QMC simulations.

To overcome this problem, we divide the parameter interval $[\lambda',\lambda'']$ into $N$ subintervals and introduce a sequence of intermediate parameters
\begin{equation}
    \lambda_0=\lambda',\quad \lambda_1,\dots,\lambda_{N-1},\quad \lambda_N=\lambda'' .
\end{equation}
The ratio of partition functions can then be decomposed as
\begin{equation}
    \frac{Z_n(\lambda'')}{Z_n(\lambda')}
=
\prod_{i=0}^{N-1}
\frac{Z_n(\lambda_{i+1})}{Z_n(\lambda_i)},
\end{equation}
where each consecutive pair of parameters $\lambda_i$ and $\lambda_{i+1}$ is chosen sufficiently close so that the corresponding probability distributions have significant overlap~\cite{Neal2001}. 
This decomposition stabilizes the numerical evaluation by ensuring that each factor can be estimated reliably within QMC simulations.

In this work, we employ the stochastic series expansion (SSE) quantum Monte Carlo method to compute the Rényi negativity $R_n$~\cite{syljusen2003directed,PhysRevE.66.046701,Evertz01012003,PhysRevB.99.165135,yan2023}. 
We choose the inverse temperature $\beta=1/T$ as the annealing parameter $\lambda$ and set the reference point at $\beta=0$, where the Rényi negativity is known to vanish. 
In the following we focus on the case $n=3$, since the cases $n=1$ and $n=2$ do not contain nontrivial entanglement information.

Within the SSE framework, the ratios of partition functions at consecutive temperatures can be estimated as
\begin{equation}
    \frac{Z_3(\beta_{i+1})}{Z_3(\beta_i)}
=
\Big\langle
\left(\frac{\beta_{i+1}}{\beta_i}\right)^{n_{\mathrm{tot}}}
\Big\rangle_{Z_3(\beta_i)},
\end{equation}
and
\begin{equation}
    \frac{Z_3^{T_B}(\beta_i)}{Z_3^{T_B}(\beta_{i+1})}
=
\Big\langle
\left(\frac{\beta_i}{\beta_{i+1}}\right)^{n_{\mathrm{tot}}}
\Big\rangle_{Z_3^{T_B}(\beta_{i+1})}.
\end{equation}
Here $n_{\mathrm{tot}}$ denotes the total number of vertex operators in the SSE operator string on the corresponding generalized path integral.

The relevant estimators are given by:$\frac{Z_{3}(\beta_{i+1})}{Z_{3}(\beta_{i})}=\langle (\frac{\beta_{i+1}}{\beta_{i}})^{n_{\text{tot}}} \rangle_{Z_{3}(\beta_{i})}$ and $\frac{Z^{T_{B}}_{3}(\beta_{i})}{Z^{T_{B}}_{3}(\beta_{i+1})}=\langle (\frac{\beta_{i}}{\beta_{i+1}})^{n_{\text{tot}}} \rangle_{Z^{T_{B}}_{3}(\beta_{i+1})}$. Here, $n_{\text{tot}}$ denotes the total number of vertex operators in the corresponding manifold: for $Z_{3}$ it refers to the non-wrapped manifold (Fig.\ref{fig:neg_def_manifold}(a)) at $\beta_{i}$, whereas for $Z^{T_{B}}_{3}$ it corresponds to the partially wrapped manifold induced by the transpose on subsystem B (Fig.\ref{fig:neg_def_manifold}(b)) at $\beta_{i+1}$. 

During the Monte Carlo sampling we record the operator count $n_{\mathrm{tot}}$ at each update step to estimate the ratios $Z_3^{T_B}(\beta_i)/Z_3^{T_B}(\beta_{i+1})$ and $Z_3(\beta_{i+1})/Z_3(\beta_i)$ for every $\beta_i$. 
The third-order Rényi negativity $R_3(\beta)$ is then obtained by accumulating these ratios along the annealing path starting from the reference point $\beta_0=0$.

\end{widetext}

\end{document}